\documentclass[pre,twocolumn,groupedaddress,floatfix]{revtex4-1}
\usepackage{amssymb,amsmath}
\usepackage{graphicx}
\usepackage{subfigure}
\usepackage[english]{babel} 
\usepackage{float}
\usepackage{color}

\usepackage{lipsum}
\begin{document}
\newcommand{\be}{\begin{equation}}
\newcommand{\ee}{\end{equation}}
\newcommand{\rojo}[1]{\textcolor{red}{#1}}

\title{Transport of localized and extended excitations in one-dimensional electrical lattices}

\author{Mario I. Molina}
\affiliation{Departamento de F\'{\i}sica, Facultad de Ciencias, Universidad de Chile, Casilla 653, Santiago, Chile}

\date{\today }

\begin{abstract} 
We study the scattering properties of a bi-inductive electrical lattice consisting of a one-dimensional array of coupled $LC$ units. For an initially localized electrical excitation, and in the absence of any impurity, we compute in closed form the mean square displacement of an initially localized electrical excitation
for the cases of an infinite and semi-infinite lattice, obtaining a ballistic propagation under very general conditions. 
For the transport of extended excitations, we compute in closed form the transmission coefficient of electro-inductive plane waves across an impurity region, containing a number of  side-coupled units, or a  single internal impurity with coupling to first-and second nearest neighbors, looking for the presence of Fano resonances (FRs). For all cases examined, we obtain a closed-form expression for the position of the FR in terms of the relative strengths of the inductive couplings involved. For the case of two, identical side-coupled impurities, the position of the single FR turns out to be independent of the relative distance between the two impurities. 

\end{abstract}

\maketitle

{\em Introduction}. The transport of localized and extended excitations inside a medium is an old, yet all-important problem in numerous areas of science and technology, many of which depend crucially on the ability to create, steer and manage the propagation of excitations. Some of these systems obey a discrete dynamics where the usual wave equations are replaced by their discrete versions. Examples of these systems include propagation of electrons in crystalline solids\cite{ashcroft,electron,electron2}, propagation of solitons in coupled waveguide arrays\cite{arrays,arrays2,arrays3,arrays4,arrays5,malomed,managment}, exciton propagation in biomolecules\cite{davidov,forster,Hu}, Bose-Einstein condensates in coupled magneto-optical traps\cite{BEC0,BEC,BEC2,BEC3}, magneto-inductive waves in magnetic metamaterials\cite{mm,mm2,mm3} and electrical waves in electrical transmission lines\cite{english,hirota,kuusela0,kuusela,english3,english4,english5}, to name some.

A periodic array of inductively coupled electrical units, constitute a highly controllable experimental testbed  in which to investigate general wave phenomena such as band structure, localized modes in the presence of disorder, and propagation of localized and extended electrical excitations. The macroscopic scale where these electrical effects take place makes electrical circuits easier to measure experimentally than in other contexts.  For a periodic electrical array, it is only natural to explore the propagation of excitations in the presence of one or few ``impurities'' that break the translational invariance. In the absence of any defects, the linear periodic electrical lattice supports the existence of electro-inductive waves. The addition of judiciously placed defects might lead to interesting resonance phenomena, such as Fano resonances (FR), where there is total reflection of plane waves through the impurity region, in an otherwise periodic potential. In a typical FR system, the wave propagation in the presence of a periodic scattering potential is characterized by closed and open channels. The open channel guides the propagating waves as long as the eigenfrequencies do not match those of the closed channels. The total reflection of waves in the open channel occurs when a localized state originating from one of the closed channels resonates with the open channel spectrum\cite{fano,fano2,fano3,fano4}. FRs have potential applications in a wide range of fields, from telecommunication to ultrasensitive biosensing, medical instrumentation, and data storage\cite{fano2}.

In this work, we study the transport of both, initially localized electrical excitations, and extended electrical excitations. In the first case, the initial position is taken at the very surface of the array, or well inside it.
In the last case, we focus on FR effects due to a few defects whose position, with respect to the periodic electrical array, can be easily tuned, making this type of configuration an interesting one to probe experimentally. More specifically, we look at the propagation of electric plane waves in a bi-inductive electrical lattice  across a localized region that contains a number of impurities that break the discrete translational invariance and look into the problem of tuning the position of these resonances in momentum space.

\begin{figure}[t]
 \includegraphics[scale=0.22]{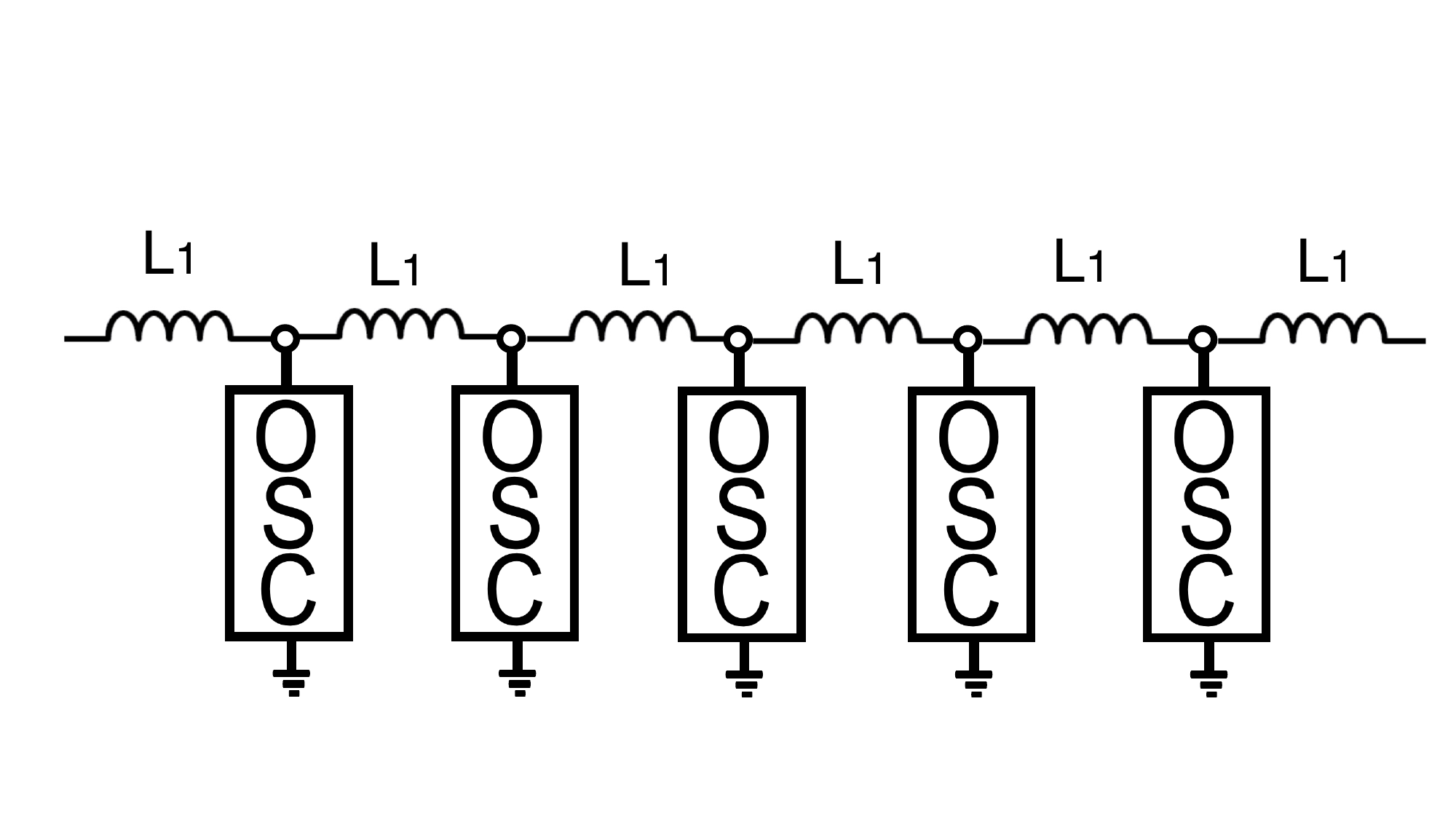}
 
 \vspace{-0.63cm}
 
  \includegraphics[scale=0.16]{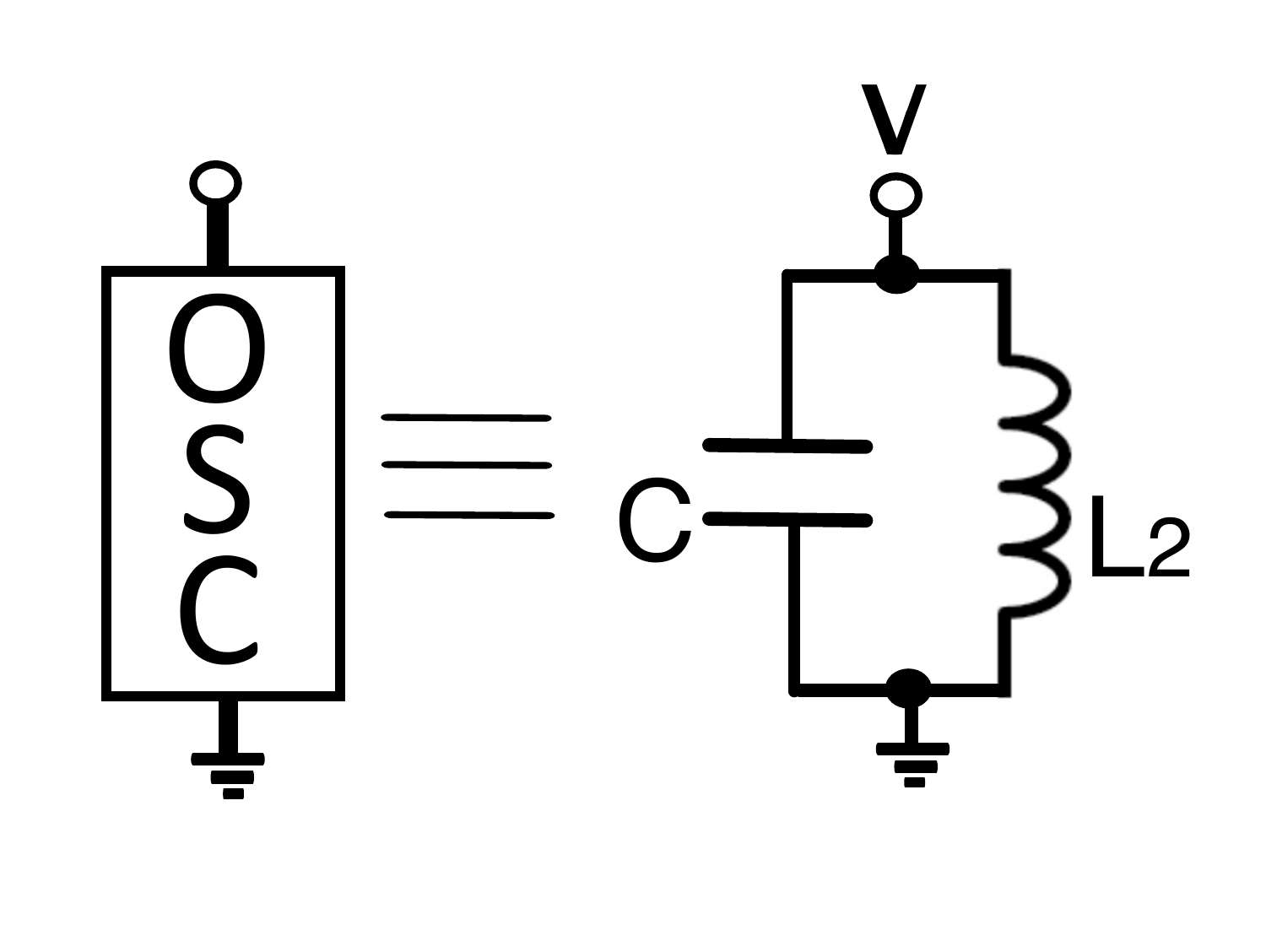}
  
  \vspace{-0.5cm}
  
  \caption{Infinite bi-inductive electrical lattice (after Shimizu et al.\cite{shimizu}).}  \label{fig1}
\end{figure}

\noindent 
{\em The model}.\ Figure 1 shows a bi-inductive electrical lattice composed of a one-dimensional array of $LC$ circuits coupled inductively. $L_{1}$ and $L_{2}$ are the inductances, $C$ is the linear capacitance, and $U_{n}$ is the voltage drop across the $n$th LC unit\cite{shimizu}. Note that each unit is an oscillating circuit characterized by a resonant frequency $\omega_{2}=1/L_{2}C$. The electrical charge $Q_{n}$ on the $n$th capacitor is given by $Q_{n} = C  U_{n}$. After the application of  Kirchhoff's law, the equations for the voltages are
\be
{d^2 Q_{n}\over{d t^2}} = {1\over{L_{1}}}(V_{n+1} - 2 V_{n} + V_{n-1}) - {1\over{L_{2}}}V_{n},\label{eq1}
\ee
where $Q_{n}=C\ V{n}$. After introducing dimensionless variables, we obtain
\be
{d^2 q_{n}\over{d \tau^2}} = q_{n+1} - 2 q_{n} + q_{n-1} - \gamma^2\ q_{n},
\label{eq2}
\ee
where $q_{n}=Q_{n}/Q_{c}$, where $Q_{c}$ is a characteristic charge, $\tau=(1/\sqrt{L_{1} C})\ t$ and $\gamma=L_{1}/L_{2}$. Note that $\gamma^2$ can also be written as $(\omega_{2}/\omega_{1})^2$ i.e., the ratio of the intra and inter resonant frequencies of the electrical array.  We look for the stationary modes, in the form $V_{n}(t) = V_{n} \cos(\Omega\ \tau + \phi)$. The stationary equation becomes
\be
-\Omega^2\ q_{n} = q_{n+1} - 2 q_{n} + q_{n-1} - \gamma^2\ q_{n}.
\ee
We look for the dispersion relation of plane waves, $q_{n} = A\ e^{i k n}$. One obtains:
\be
\Omega_k^2 = 4 \sin(k/2)^2 + \gamma^2.\label{dispersion}
\ee

\noindent
{\em Propagation of initially localized electrical excitations}. Let us 
consider the general problem as computing the propagation of an electrical excitation that is  initially completely localized on one of the units, along a homogeneous array. Let us take a generic dispersion $\Omega_k^2$ that satisfies $\Omega_{-k}^2 = \Omega_{k}^2$ (like in (\ref{dispersion}) ). A useful observable to monitor the excitation transport for this problem is the mean square displacement (MSD). The MSD is defined as
\be
\langle n^2 \rangle = \sum_{n} n^2 |q_{n}(\tau)|^2 / \sum_{n} |q_{n}(\tau)|^2
\label{MSD}
\ee
Let us look at the MSD for a completely localized initial charge on a capacitor,  $q_{n}(0)=A\ \delta_{n 0}$ and no currents initially, $(d q_{n}/d \tau)(0)=0$. We have two cases of interest, an infinite lattice and a semi-infinite lattice.\\

(a) Infinite lattice: we have formally
\begin{eqnarray*}
 q_{n}(\tau) &=& (A/4 \pi) \int_{-\pi}^{\pi} e^{i (k n-\Omega_{k})\tau} dk\\
            & & + (A/4 \pi) \int_{-\pi}^{\pi} e^{i (k n+\Omega_{k})\tau} dk
\end{eqnarray*}
where $\Omega_{k}$ is the dispersion.
 After replacing this form for $q_{n}(\tau)$ into Eq.(\ref{MSD}), one obtains after some algebra, a closed form expression for $\langle n^2 \rangle$:
\be
\langle n^2 \rangle = {(1/2\pi)\ \int_{-\pi}^{\pi}d k (d \Omega_{k}/d k)^2 (1 - \cos(2\ \Omega_{k}\ \tau))\ \tau^2 
\over{1 + (1/2\pi) \int_{-\pi}^{\pi} d k\ \cos(2\ \Omega_{k}\ \tau)}}.\label{n2closed}
\ee
As time $\tau$ increases, the contributions from the cosine terms to the integrals decrease and, at long times, $\langle n^2\rangle$ approaches a ballistic behavior: 
\be
\langle n^2 \rangle \approx \left[ {1\over{2 \pi}} \int_{-\pi}^{\pi} \left( {d \Omega(k)\over{d k}}\right)^2\ dk\right]\ \tau^2 \hspace{1cm} (\tau\rightarrow \infty).\label{ballistic1}
\ee
For the special case of our dispersion (\ref{dispersion}), we obtain
\be
\langle n^2 \rangle \approx (1/4) \{\ 2 + \gamma (\gamma - \sqrt{4 + \gamma^2})\ \}\ \tau^2 \equiv v^2 \tau^2\label{eq8}
\ee  
where $v$ plays the role of a characteristic speed. 
At short times,
\be
\langle n^2 \rangle \approx \left[ {1\over{2 \pi}} \int_{-\pi}^{\pi} \left( \Omega_{k}{d \Omega_{k}\over{d k}} \right)^2 dk\right] \ \tau^4 \hspace{1cm}(\tau\rightarrow 0),
\ee 
for our case (\ref{dispersion}) this implies,
\be
\langle n^2 \rangle \approx (1/2)\ \tau^4\hspace{1cm}(\tau\rightarrow 0).
\ee 

(b) Semi-Infinite lattice: This case is more complex than the previous one because  now we must take into account the presence of the boundary at $n=0$. The way to solve this problem is to use the method of images: Because $q_{n}=0$ to the left of $n=0$, we impose that $q_{-1}=0$. In terms of the solution for the infinite lattice $q_{n}^{\infty}$, this implies:
$q_{n}(\tau)= q_{n}^{\infty} - q_{-n-2}^{\infty}$. Thus, for a completely localized charge excitation at $n=0$ (and its accompanying image at $n=-2$) and with no currents present initially, we have
\begin{eqnarray*}
\lefteqn{ q_{n}(\tau) = 
 (A/4 \pi) \int_{-\pi}^{\pi} e^{i (k n-\Omega_{k}\tau)} dk}\\
            & & + (A/4 \pi) \int_{-\pi}^{\pi} e^{i (k n+\Omega_{k}\tau)} dk - (A/4 \pi) \int_{-\pi}^{\pi} e^{-i (k (n+2)+\Omega_{k}\tau)} dk\\ & &
-(A/4 \pi) \int_{-\pi}^{\pi} e^{i (-k (n+2)+\Omega_{k}\tau)} dk            
\end{eqnarray*}
After replacing this form into Eq.({MSD}) we obtain, after some lengthy algebra:
\be
\langle n^2 \rangle = {\int_{-\pi}^{\pi}\{ n_{1}(k)+n_{2}(k)+n_{3}(k)+n_{4}(k)+n_{5}(k) \}\ dk\over{\int_{-\pi}^{\pi}\{ d_{1}(k)+d_{2}(k) \}}}
\ee
where,
\be
n_{1}(k)=2 \left({d \Omega_{k}\over{d k}}  \right)^2 \tau^2 +
2 \left( 2+ \left({d \Omega_{k}\over{d k}}\right) \tau \right)^2,
\ee
\be
n_{2}(k)=\left( 8 - 2 \left({d \Omega_{k}\over{d k}} \right)^2 \tau^2 \right) \cos(2 \Omega_{k} \tau), 
\ee
\be
n_{3}(k)=4 \left({d \Omega_{k}\over{d k}}  \right)^2 \cos(2 k) \ \tau^2,
\ee  
\be
n_{4}(k)= -2 \tau \left({d \Omega_{k}\over{d k}}  \right)\left(2+\left({d \Omega_{k}\over{d k}}  \right) \tau\right) \cos(2 k + 2 \Omega_{k} \tau),
\ee
\be
n_{5}(k)= 2 \tau \left({d \Omega_{k}\over{d k}}  \right)\left(2-\tau \left({d \Omega_{k}\over{d k}}\right)\right) \cos(2 k - 2\Omega_{k} \tau),
\ee
\be
d_{1}(k)=  4 (1+\cos(2 \Omega_{k} \tau)),
\ee
\be
d_{2}(k)=-4 \cos(2 k + 2 \Omega_{k} \tau).
\ee  
At long times, the MSD reduces to
\begin{eqnarray}
\langle n^2 \rangle & & \approx \left[ {1\over{2 \pi}} \int_{-\pi}^{\pi} \left( {d \Omega(k)\over{d k}}\right)^2 (1+ \cos(2 k))\ dk\right]\ \tau^2\nonumber\ \ \ \ \ \ \\
& & \hspace{4.5cm}(\tau\rightarrow \infty).\label{ballistic1}
\end{eqnarray}
 i.e., a ballistic propagation.  For our dispersion (\ref{dispersion}), we obtain:
\begin{eqnarray}
\langle n^2 \rangle & & \approx (1/8) (2 + \gamma^2) (2 + 
   \gamma (-2 \sqrt{4 + \gamma^2} + \\
& & \gamma (4 + \gamma^2 - \gamma \sqrt{4 + \gamma^2})))\ \tau^2
\hspace{0.5cm}(\tau\rightarrow \infty)
\end{eqnarray}
while at short times,
\be 
\langle n^2 \rangle \approx 2 \hspace{0.9cm}(\tau\rightarrow 0)
\ee
Result (\ref{ballistic1}) is valid for a broad class of discrete periodic systems characterized by a dispersion $\Omega_{k}$ that obeys $\Omega_{-k}=\Omega_{k}$.  Figure 2 shows both, the MSD for the infinite and semi-infinite lattice, and their characteristic ballistic `speed'.
\begin{figure}[t]
 \includegraphics[scale=0.275]{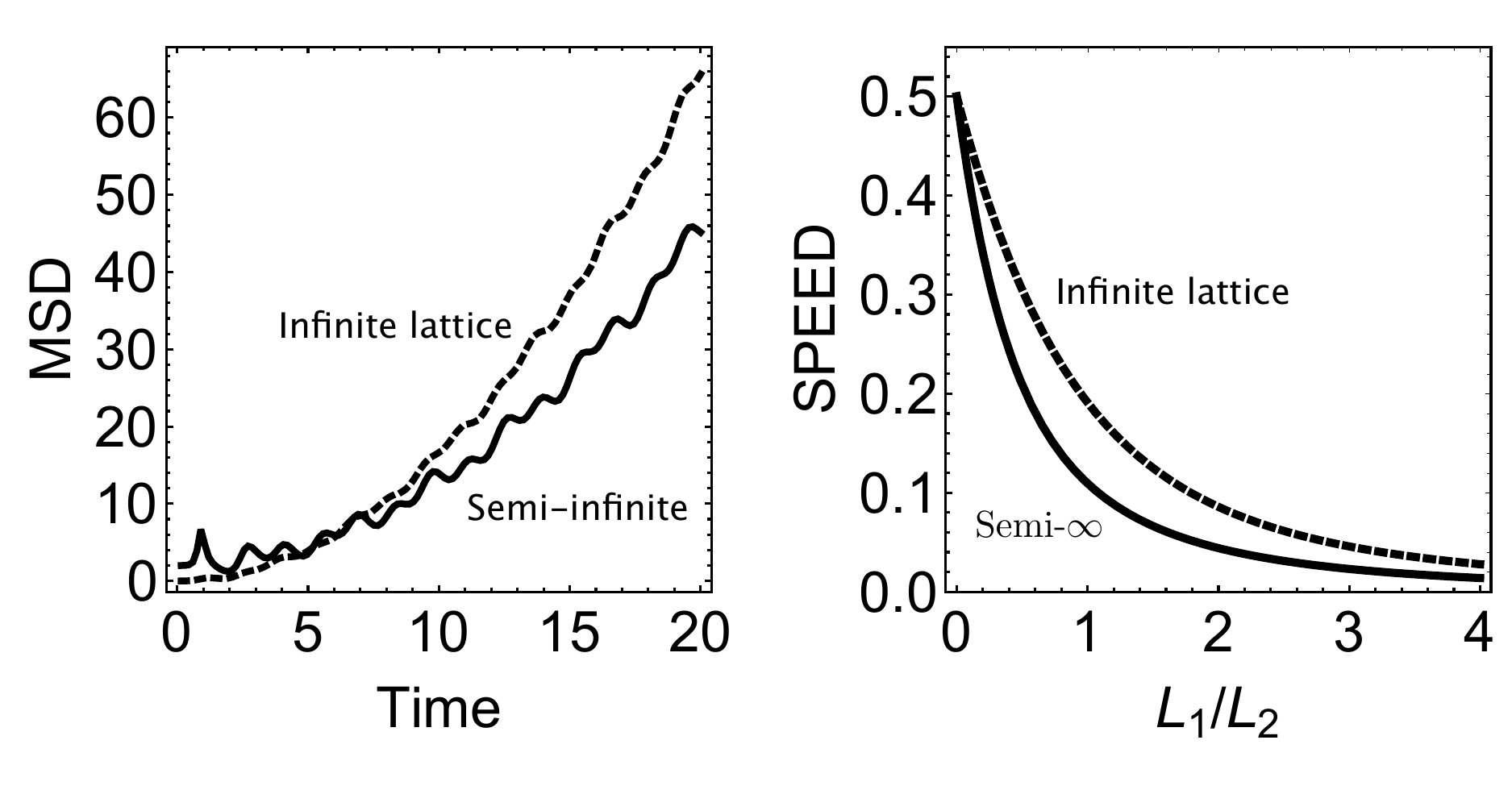}
 \caption{Left: Mean square displacement (MSD) vs time for an infinite (dashed) and semi-infinite (solid) electric lattice. Right: Ballistic speed  vs the ratio of inductances for an infinite (dashed) and semi-infinite (solid) electric lattice.}  \label{fig2}
\end{figure}

\noindent
{\em Propagation of extended electrical excitations}. Let us now consider the transmission of plane waves across a finite segment that contains a number of  impurities, looking for the presence of Fano resonances (FRs), typified by a complete reflection of plane waves.  We will consider several cases:\\
(a) Single capacitive impurity inside an array with coupling to first and second nearest neighbors (Fig.3). 
\begin{figure}[t]
 \includegraphics[scale=0.273]{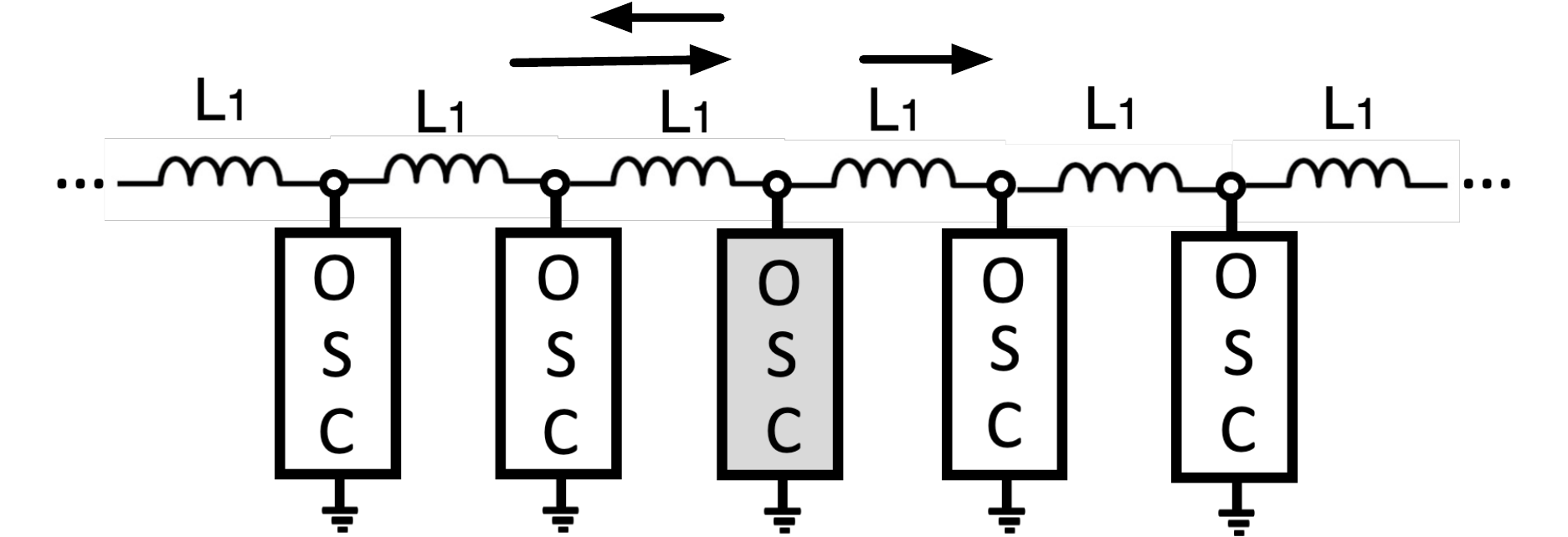}
 \includegraphics[scale=0.23]{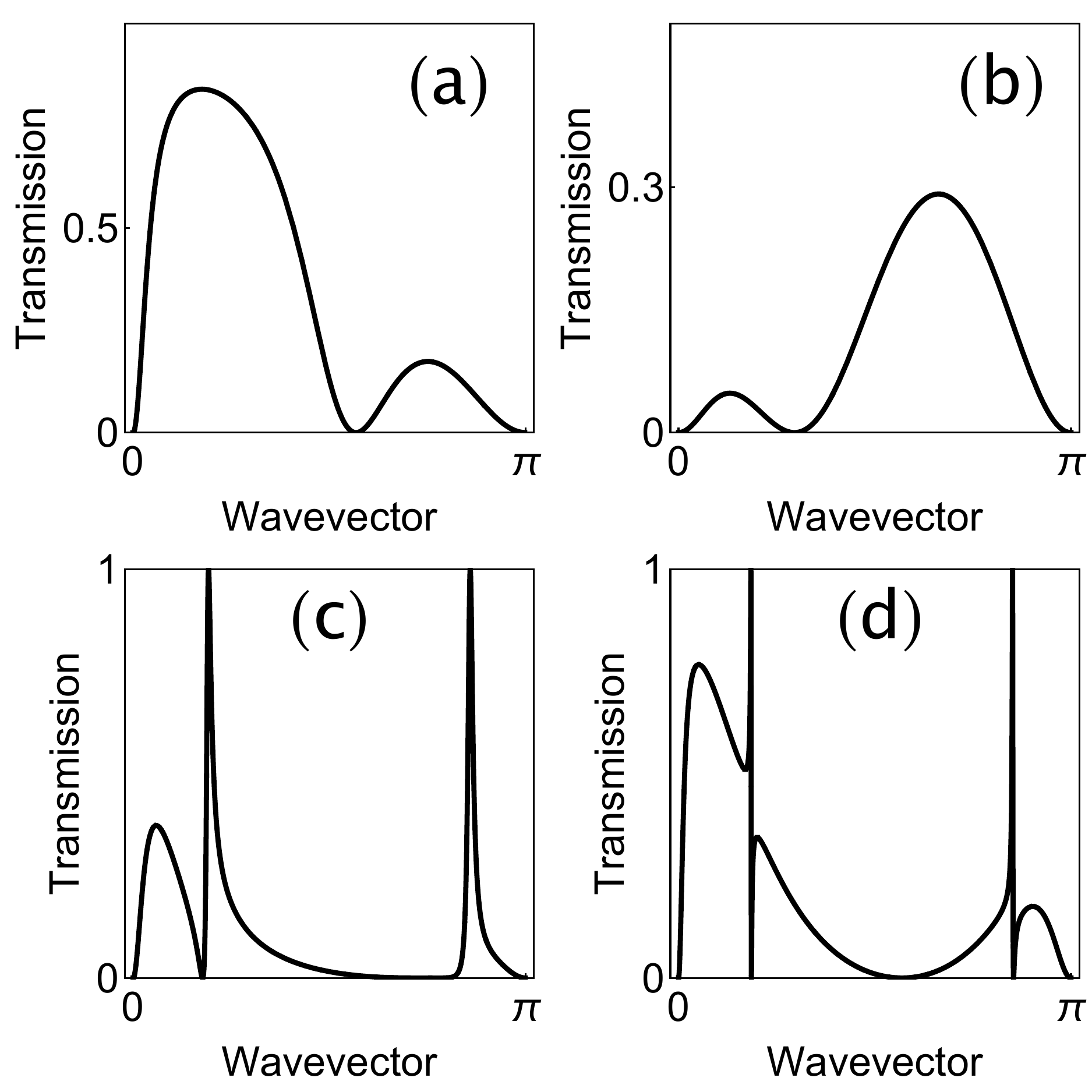}
 \vspace{0cm}
 \caption{ Bi-inductive electrical lattice containing a single capacitive, in an array with coupling to first-and second nearest neighbors. (a) $\gamma'=0.5, \Delta=0.5$, (b) $\gamma'=0.5, \Delta=3.2$, (c) $\gamma'=2.5, \Delta=4$, (d) $\gamma'=3, \Delta=2$.  }
\end{figure}
The inductive coupling originates from a dipole-dipole interaction, which decays slowly in space like $1/d^3$ where $d$ is the distance between two units. Thus, the presence of a second nearest neighbor interaction is not farfetched. The capacitance of the units is $C$, while at the impurity site ($n=0$) its value is $C_{0}$. 
The equations are
\begin{eqnarray}
-\Omega^2 q_{n} &=& q_{n+1}-2 q_{n}+q_{n-1} + \gamma'^2 (q_{n+2}-2 q_{n}+q_{n-2})\nonumber\\
& & -\gamma^2 q_{n}+\Delta \Omega^2 q_{n} \delta_{n 0}  
\end{eqnarray}
where, $\Delta=(c_{0}-c)/c$ is the capacitance mismatch, $\gamma^2=L_{1}/L_{2}$ and $\gamma'^2=L_{1}/L'$ is the inductive coupling to second nearest neighbors. Assuming a plane wave solution of the form
\begin{figure}[t]
 \includegraphics[scale=0.275]{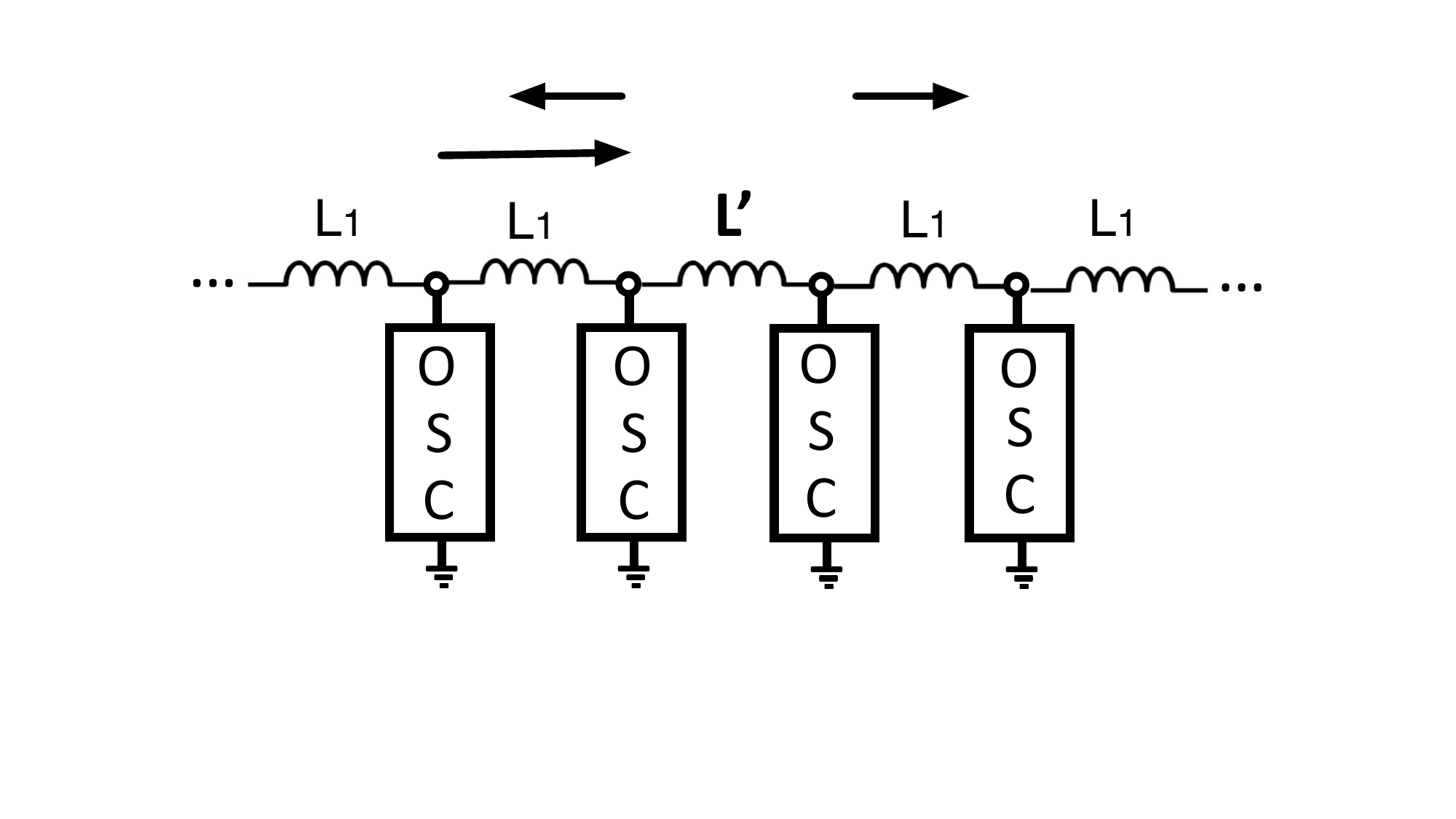}
 \vspace{-1cm}
 \caption{Bi-inductive electrical lattice containing a single coupling impurity. }
\end{figure}
\be
q_{n}(t)=\left\{ \begin{array}{ll}
			A\ e^{i k n} + B\ e^{-i k n} & \mbox{$n<0$}\\
			T\ e^{i k n} &\mbox{$n\ge 0$}
			\end{array}
				\right.\label{solution}
\ee
we obtain a closed-form expression for the transmission coefficient $t \equiv |T/A|^2$:
\be
t = \left|{\alpha(k)\over{\beta(k) + \delta(k) + \epsilon(k)+\eta(k)+\mu(k)}}\right|^2
\ee
with\\
\begin{eqnarray}
\alpha(k)&=& 2(1+3 \gamma'^4-\Delta \gamma'^2 (2+\gamma^2+2 \gamma'^2)+\nonumber\\
 & &+2(3+\Delta)(\cos(k)+\gamma'^2 \cos(2 k))+\nonumber\\
 & &+ 2 \gamma'^6 \cos(3 k))\sin(k)\nonumber\\
\beta(k)&=&3(1+\Delta)e^{i k}\gamma'^4+(\Delta-3)e^{7 i k}\gamma'^4+(1+\Delta)\gamma'^6\nonumber\\
\delta(k)&=& (\Delta-1)e^{8 i k}\gamma'^6+e^{3 i k}(1+\Delta-2 \Delta (2+\gamma^2)\gamma'^2+\nonumber\\
& & -4 \Delta \gamma'^4)\nonumber\\
\epsilon(k)&=& e^{5 i k}(-1+\Delta+2 \gamma'^4\Delta)-\nonumber\\
& & e^{6 i k}\gamma'^2(3-\Delta+\Delta(2+\gamma^2))\gamma'^2+(3\Delta-1)\gamma'^4\nonumber\\
& & \eta(k)=\Delta e^{4 i k}(-2-\gamma^2+(2+\gamma^2)\gamma'^4+4 \gamma'^6)\nonumber\\
\mu(k)&=&-e^{2 i k}\gamma'^2 (-3+\gamma'^4+\Delta(-3+(2+\gamma^2)\gamma'^2+\nonumber\\ 
& & 3 \gamma'^4))
\end{eqnarray}
Figure 3 shows some examples of transmission curves for various $\Delta$ and $\gamma'^2$ values. We see that the system can support the presence of one, two, and three FRs. When the coupling to second nearest neighbors is taken as zero, $\gamma'=0$, no FRs are present. 

Another sub-case is to have only the capacitive impurity endowed with coupling to first-and second nearest neighbors. In this case, it can be proved (not shown) that there is a single FR at $k=\arccos(-1/2 \gamma'^2)$, provided $\gamma'>1/\sqrt{2}$. 

(b) Coupling impurity: In this case we introduce a different coupling between two given electrical units only (Fig.4). This is achieved by means of an inductor $L'$ between say, sites $0$ and $1$. The equations are
\begin{eqnarray}
& & \Omega^2 q_{n}+(q_{n+1}-2 q_{n} + q_{n-1}) - \gamma^2 q_{n}+
\nonumber\\
& & (q_{0}-q_{1})(1-\gamma'^2)(\delta_{n 0}-\delta_{n 1})=0 
\end{eqnarray}

After posing a solution in the form (\ref{solution}), one obtains
\be
t = \left| {\gamma'^2 (1 + e^{i k})\over{1+(2 \gamma'^2-1)e^{i k}}}  \right|^2
\ee
which is zero at $k=\pi$ only. Thus, no FR in this case. This case is reminiscent of the case of a single site impurity that does not have a FR, too. It would seem that in these two cases, the system does not possess  enough internal structure to bring about the necessary interference for a FR to occur. 

\noindent
(c) The next case we examine consists of a single side-coupled impurity where one of the electrical units is coupled to a single unit in the chain, with inductive coupling $L'$ (Fig.5). This geometrical configuration is one commonly used in studies of FRs. The equations are
\begin{eqnarray}
-\Omega^2 q_{n}&=&(q_{n+1}-2 q_{n}+q_{n-1}) - \gamma^2 q_{n} + 
\gamma'^2 (q_{e}-q_{0})\delta_{n 0}\nonumber\\
-\Omega^2 q_{e} &=& \gamma'^2 (q_{0}-q_{e})-\gamma^2 q_{e} \label{side_defect}
\end{eqnarray}
\begin{figure}[t]
 \includegraphics[scale=0.273]{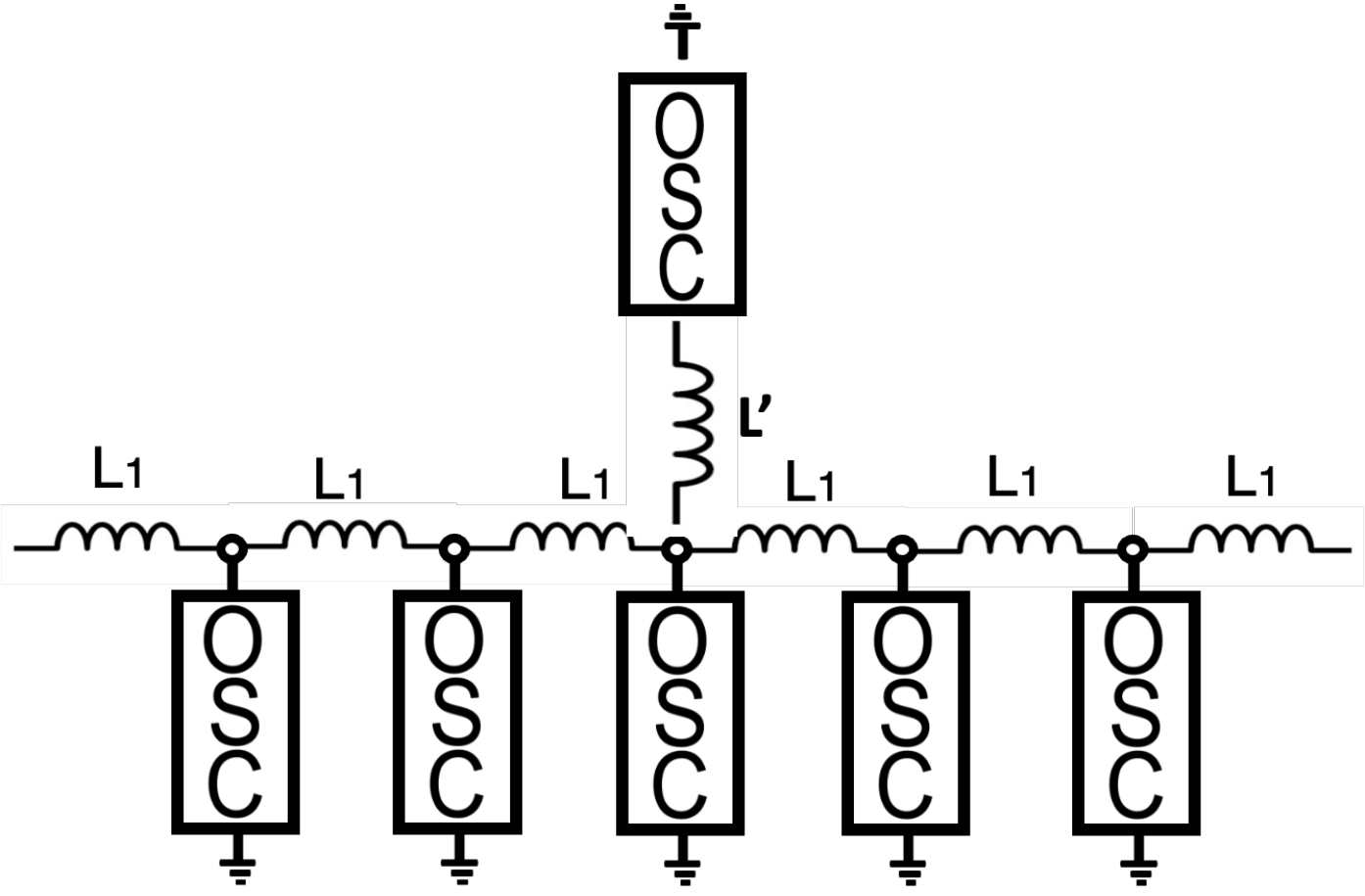}
  \includegraphics[scale=0.23]{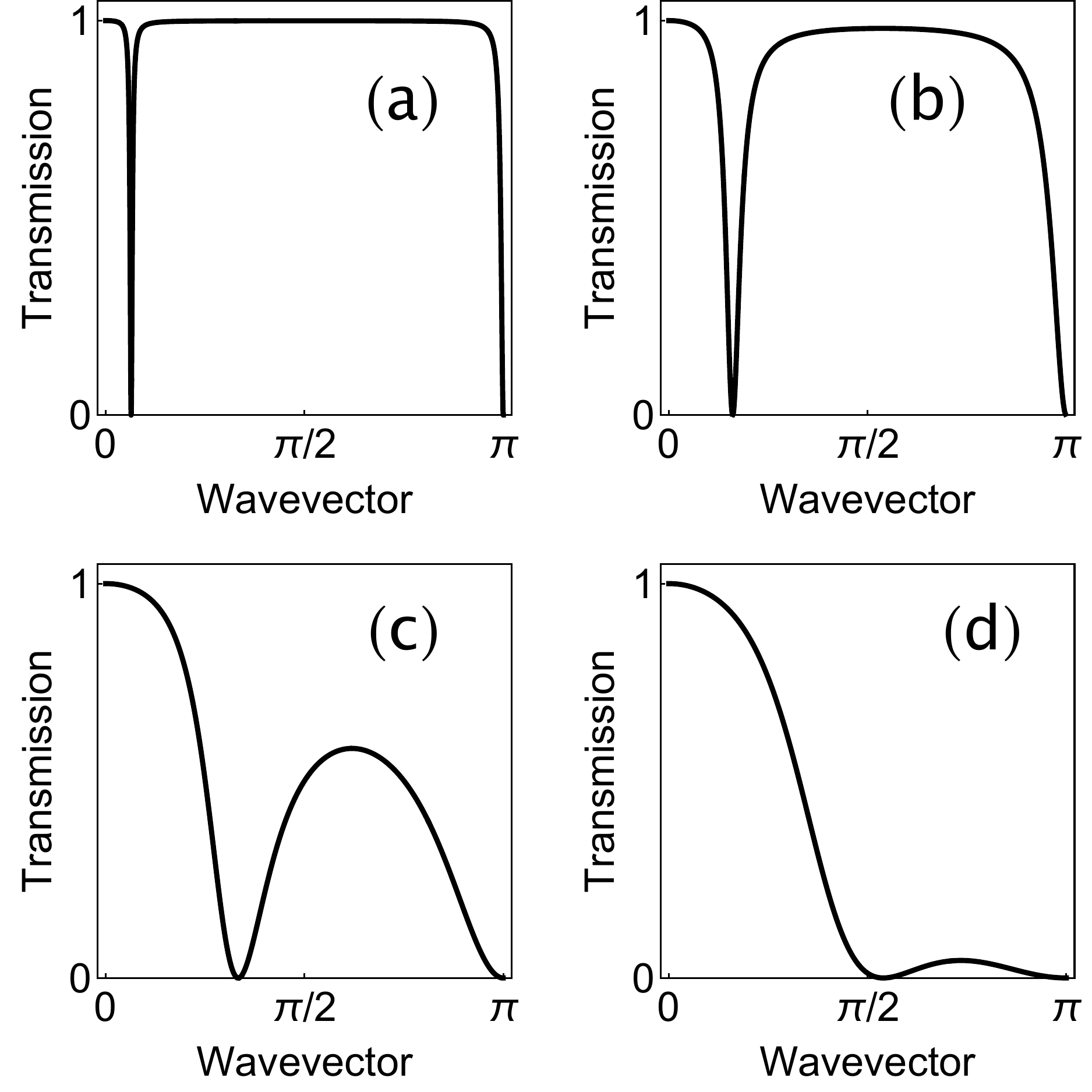}
 \caption{Top: Bi-inductive electrical lattice containing a single side-coupled impurity. Bottom: Transmission coefficient of plane waves vs wavevector for (a) $\gamma'=0.2$, (b) $\gamma'=0.5$,(c) $\gamma'=1.0$ and (d) $\gamma'=1.5$}.
\end{figure}
After posing the plane wave ansatz, we obtain
\be
t = \left|{e^{i k}(1+e^{i k})(-2+\gamma'^2+2 \cos(k)) \over{1-e^{2 i k}+e^{3 i k}+e^{i k}(-1+2 \gamma'^2)}}\right|^2
\ee
This means that a single FR at 
\be  
k = \arccos(1-(\gamma'^2/2)))
\ee
exists, provided $\gamma'<2$.
Thus, the FR can be tuned to occur inside $0<k<\pi$ if we sweep
$\gamma'$ between $\gamma'=0$ and $\gamma'=2$ or $0<\sqrt{L_{1}/L'}<4$.

\noindent
(d) Finally, let us consider the FR problem for the case of two identical lateral defects, separated by a distance of $d$ units (Fig.6). The charges on these side 
\begin{figure}[t]
 \includegraphics[scale=0.3]{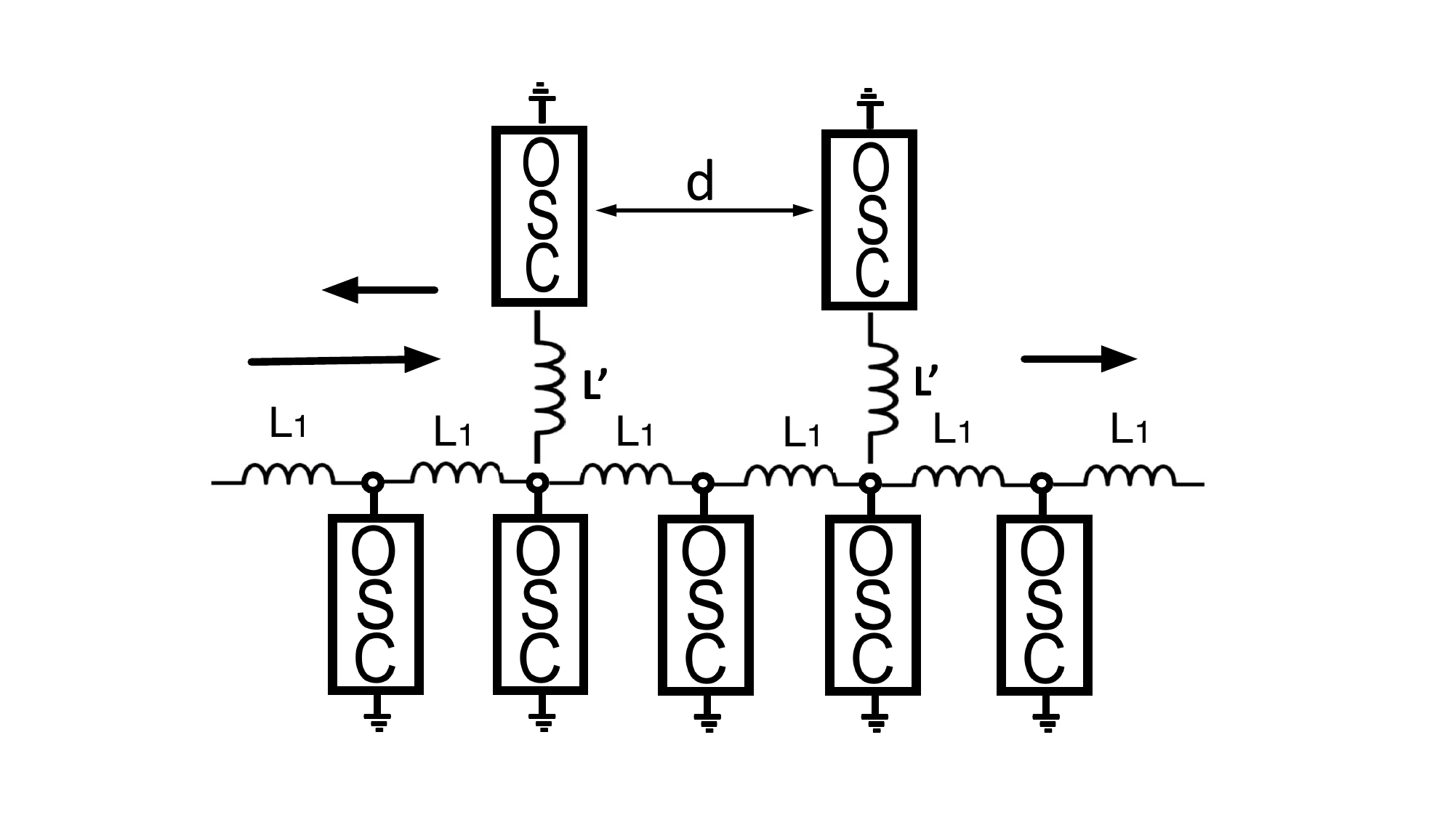}

  \includegraphics[scale=0.25]{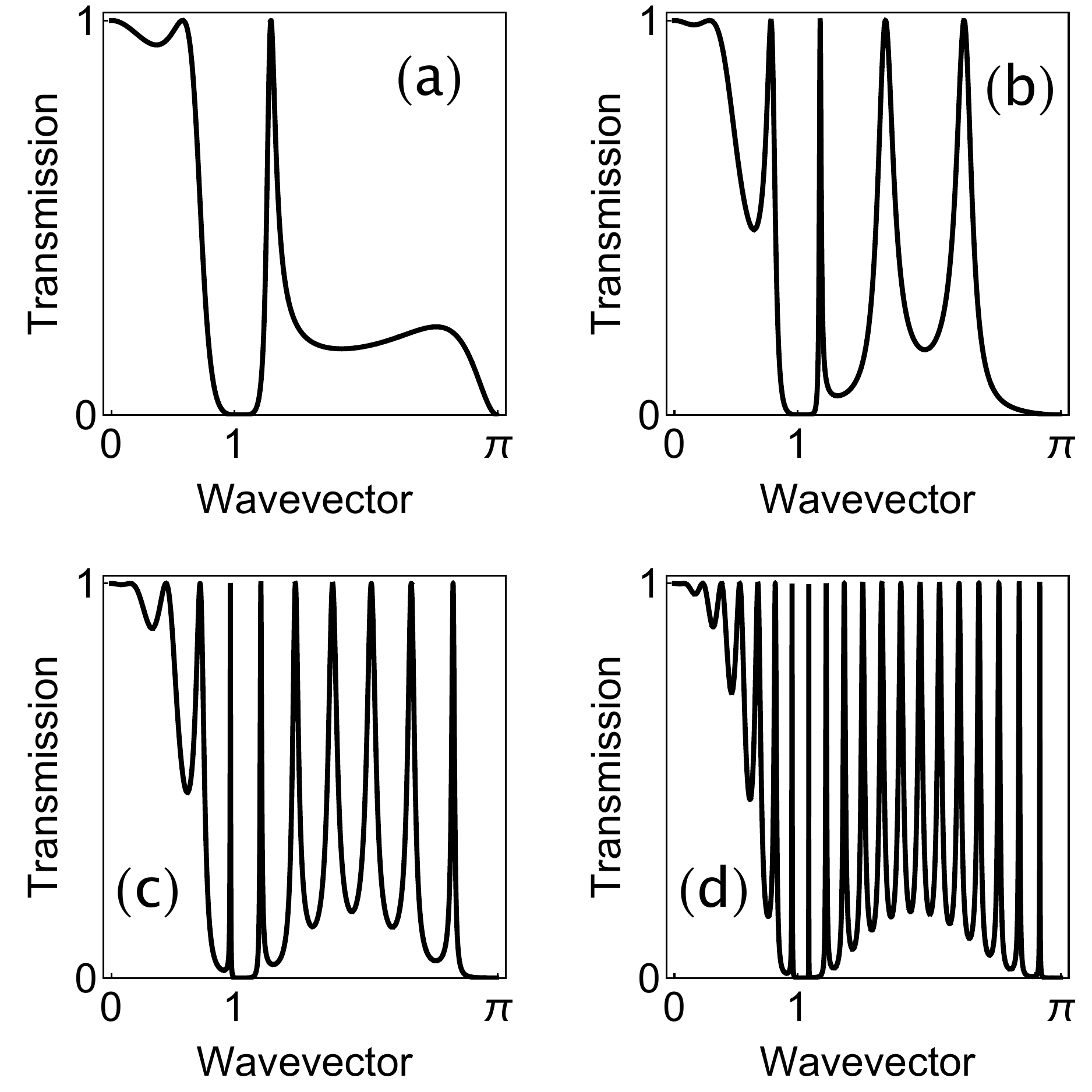}
 \caption{Top: Bi-inductive electrical lattice containing two side-coupled impurities separated by a distance of $d$ units. Bottom: Transmission coefficient vs wavevector for $\gamma'=1$ and (a) $d=2$, (b) $d=5$, (c) $d=10$ and (d) $d=20$.}
\end{figure}
units are denoted as $q_{A}$ and $q_{B}$. The equations are
\begin{eqnarray}
& &\Omega^2+(q_{n+1}-2 q_{n}+q_{n-1})-\gamma^2 q_{n}+\gamma'^2 (q_{A}-q_{0})\delta_{n 0}+\nonumber\\
& & +\gamma'^2(q_{B}-q_{d})\delta_{n d}=0\nonumber\\
& & \Omega^2 q_{A} + \gamma'^2 (q_{0}-q_{A}) - \gamma^2 q_{A}=0\nonumber\\
& & \Omega^2 q_{B}+\gamma'^2 (q_{1}-q_{B})-\gamma^2 q_{B}=0.
\end{eqnarray}
As before, we pose a plane-wave solution
\be
q_{n}(t)=\left\{ \begin{array}{ll}
			A\ e^{i k n} + B\ e^{-i k n} & \mbox{$n<0$}\\
			C \ e^{i k n} + D e^{-i k n} & \mbox{$0<n<d$}\\
			T\ e^{i k n} &\mbox{$n\ge d$}
			\end{array}
				\right.\label{solution2}
\ee
For a given $d$ value, a closed-form expression can be obtained for the transmission. For instance, for a separation $d=20$, one obtains
\be
t = \left| {\alpha(k)\over{\beta(k) + \delta(k) + \eta(k)}} \right|^2
\ee
where,
\begin{eqnarray}
\alpha(k)&=&(1+e^{i k})(-2+\gamma'^2+2 \cos(k))^2\nonumber\\
\beta(k)&=& -1-4\gamma'^4\sum_{n=6}^{41}(-1)^{n}e^{i k n}-3 \gamma'^4e^{42 i k}+\gamma'^4 e^{43 i k}\nonumber\\
\delta(k)&=&(3-4\gamma'^4)e^{5 i k}+(3-4 \gamma'^4)e^{4 i k}+\nonumber\\
& & e^{2 i k}(-2+8\gamma'^2-4\gamma'^4)\nonumber\\
\eta(k)&=& (-1+4 \gamma'^4)e^{5 i k}+ (-2-4\gamma'^2+4 \gamma'^4)e^{3 i k}
\end{eqnarray}
Results are shown in Fig.6. A FR at wavevector $k$ (other than $\pi$) will occur whenever $t=0$, that is,
\be
k = \arccos(1-(\gamma'^2/2))\label{k}
\ee
which is well-defined provided $\gamma^2 < 2$. As suggested by Fig.6, 
result (\ref{k}) is actually valid for any $d$, as computer calculations show.

\noindent
{\em Conclusions}. We have investigated the propagation of localized and extended electrical excitations in an electric lattice composed of an array of $LC$ units coupled inductively. For the case of the localized excitation in a homogeneous array, we found the mean square displacement of an initially excited LC unit, as a function of time, in closed form. At long times it was shown that the propagation is ballistic. This result is actually valid for a wide family of systems described by a dispersion relation satisfying very general conditions.  For extended excitations represented by a set of electrical plane waves, we computed the transmission coefficient across a finite impurity region and their associate FR, in closed form. For both, the single capacitance impurity and the single inductive impurity no FRs were found, while for the case of a single capacitance impurity in an array with couplings to first -and second nearest neighbors up to three FRs are possible. Interestingly, for the case of a singe and double side-coupled impurities, only a single FR was found, independently of the distance along the array between the two inductive impurities.

Experimental observation of these resonances, could in principle be achieved by means of an `electric plane wave' propagating along the LC array. This plane wave would consist of a broad gaussian electrical pulse launched at the beginning of the LC array with a given momentum $k$. Now, since this implies a large number of electrical units, it is perhaps more realistically to use printed microscopic circuitry\cite{nanoscopic} that should incorporate a way to compensate for radiative and Ohmmic losses, such as tunnel (Esaki) diodes\cite{esaki1,esaki2}.

\acknowledgments
This work was supported by Fondecyt Grant 1200120.


\begin{thebibliography}{99}

\bibitem{ashcroft}
N. W. Ashcroft and N. D. Mermin, Solid State Physics (Thomson Learning, Toronto, 1976).

\bibitem{electron}
Sindhunil Barman Roy, Electrons in crystalline solids, in Mott Insulators, IOP Publishing Ltd 2019.

\bibitem{electron2}
C. M. Goringe, D. R. Bowler, and E. Hernandez, Tight-binding modeling of materials.
Rep. Prog. Phys. {\bf 60}, 1447-1512 (1997).

\bibitem{mm}
E Shamonina and L Solymar, Magneto-inductive waves supported by
metamaterial elements: components for a one-dimensional waveguide,
J. Phys. D: Appl. Phys. {\bf 37}, 362-367 (2004).

\bibitem{mm2}
O. Sydoruk, O. Zhuromskyy, A. Radkovskaya, E. Shamonina, L. Solymar, 
Magnetoinductive Waves I: Theory in  Theory and Phenomena of Metamaterials, 
F. Capolino(Ed.),  CRC Press 2009. 

\bibitem{mm3}
M I. Molina, Defect Modes, Fano Resonances
and Embedded States in Magnetic Metamaterials,  
in Progress Optical Sci., Photonics 277-307 (2013), Springer-Verlag Berlin Heidelberg 2012.

\bibitem{arrays}
Falk Lederer, George I. Stegeman, Demetri N. Christodoulides, Gaetano Assanto, Moti Segev, and Yaron Silberberg, Discrete solitons in optics, Phys. Rep. {\bf 463}, 1-126 (2008).

\bibitem{arrays2}
H. S. Eisenberg, Y. Silberberg, R. Morandotti, A. R. Boyd, and J. S. Aitchison, 
Discrete Spatial Optical Solitons in Waveguide Arrays, Phys. Rev. Lett. {\bf 81}, 3383 (1998).

\bibitem{arrays3}
A. Sukhorukov, Y. S. Kivshar, H. S. Eisenberg, and Y. Silberberg, Spatial Optical Solitons in Waveguide Arrays, IEEE J. Quantum Elec. {\bf 39}, 31 (2003).

\bibitem{arrays4}
Demetrios N. Christodoulides, Falk Lederer and Yaron Silberberg , 
Discretizing light behaviour in linear and nonlinear waveguide lattices,
Nature {\bf 424}, 817-823 (2003).

\bibitem{arrays5}
J.W. Fleischer, M. Segev, N.K. Efremidis, and D.N. Christodoulides, 
Observation of two-dimensional discrete solitons in optically-induced nonlinear photonic lattices, Nature {\bf 422}, 147-150 (2003).

\bibitem{malomed}
B. A. Malomed, Nonlinearity and Discreteness: Solitons in Lattices. In: P. Kevrekidis, J. Cuevas-Maraver, A. Saxena (eds) Emerging Frontiers in Nonlinear Science. Nonlinear Systems and Complexity, Springer 2020.

\bibitem{managment}
 H.S. Eisenberg, Y. Silberberg, R. Morandotti, J.S.
Aitchison, Diffraction Management, Phys. Rev. Lett. {\bf 85}, 1863 (2000).

\bibitem{davidov}
A.S. Davydov, Biology and Quantum Mechanics, Pergamon Press, Oxford, 1982.

\bibitem{forster}
T. Forster, `Delocalized excitation and excitation transfer', Modern 
Quantum Chemistry, Istanbul Lectures, vol. 3, p. 93, 1965.

\bibitem{Hu}
X. Hu and K. Schulten, `How nature harvests sunlight', Physics Today {\bf 50}, 28 (1997).

\bibitem{BEC0}
Wolfgang Ketterle, Experimental Studies of Bose-Einstein Condensation,
Physics Today {\bf 52}, 30 (1999).

\bibitem{BEC}
A. Trombettoni, A. Smerzi, Discrete Solitons and Breathers with Dilute Bose-Einstein Condensates, Phys. Rev. Lett. {\bf 86}, 2353 (2001).

\bibitem{BEC2}
Chaudhuri, Saptarishi, et al. Bose-Einstein Condensation in Optical Traps and in a 1D Optical Lattice, Current Science {\bf 95}, 1026 (2008).

\bibitem{BEC3}
B. Eiermann, Th. Anker, M. Albiez, M. Taglieber, P. Treutlein, K.-P. Marzlin, and M. K. Oberthaler, Bright Bose-Einstein Gap Solitons of Atoms with Repulsive Interaction, Phys. Rev. Lett. {\bf 92}, 230401 (2004).

\bibitem{english}
M. Remoissenet, Linear Waves in Electrical Transmission Lines. In: Waves Called Solitons. Springer, Berlin, Heidelberg (1996).

\bibitem{hirota}
R. Hirota and K. Suzuki, Studies on Lattice Solitons by Using Electrical Networks, 
J. Phys. Soc, Japan {\bf 28}, 1366 (1970).

\bibitem{kuusela0}
T.Kuusela, Soliton experiments in a damped ac-driven nonlinear electrical transmission line {\bf 167}, 54 (1992).

\bibitem{kuusela}
T. Kuusela, J. Hietarinta and B. A. Malomed, Numerical study of solitons in the damped AC-driven Toda lattice, J. Phys. A {\bf 26}, L21 (1993). 

\bibitem{english3}
Ryan Stearrett, L.Q. English, Experimental Generation of Intrinsic Localized Modes in a discrete electrical transmission line, J. Phys. D {\bf 40}, 5394 (2007).

\bibitem{english4}
L.Q. English, F. Palmero, J.F. Stormes, J. Cuevas, Carretero-Gonz\'{a}lez, P. G. Kevrekidis, Nonlinear localized modes in two-dimensional electrical lattices, 
Phys. Rev. E {\bf 88}, 022912 (2013).

\bibitem{english5}
M. I. Molina, L. Q. English, Ming-Hua Chang, and P. G. Kevrekidis, Linear impurity modes in an electrical lattice: Theory and experiment, Phys. Rev. E {\bf 100}, 062114 (2019).

\bibitem{fano}
For a recent review see, for instance, A. E. Miroschnichenko, S. Flach, and Y. S. Kivshar, Fano resonances in nanoscale structures, Rev. Mod. Phys. 82, 2257 (2010).

\bibitem{fano2}
E. Kamenetskii, A. Sadreev, A.  Miroshnichenko (Eds.), Fano Resonances in Optics and Microwaves, Springer 2018.

\bibitem{fano3}
Ajith Ramachandran, Carlo Danieli, and Sergej Flach, Fano Resonances in Flat Band Networks, in: E. Kamenetskii, A. Sadreev, A. Miroshnichenko (eds) Fano Resonances in Optics and Microwaves, Springer 2018.

\bibitem{fano4}
U. Naether and M. I. Molina, Fano resonances in magnetic metamaterials, Phys. Rev. A 84, 043808 (2011).

\bibitem{shimizu}
Kuniyasu Shimizu, Tetsuro Endo, and Daishin Ueyama, Pulse Wave Propagation in a Large Number of Coupled Bistable Oscillators, IEICE Trans. Fundamentals {\bf E91}, 
2540 (2008)
\bibitem{nanoscopic}
N. Engheta, A. Salandrino, and A. Alu, Circuit Elements
at Optical Frequencies: Nanoinductors, Nanocapacitors,
and Nanoresistors, Phys. Rev. Lett. {\bf 95}, 095504 (2005).

\bibitem{esaki1}
L. Esaki, New Phenomenon in Narrow Germanium $p-n$
Junctions,  Phys. Rev. 109, 603 (1958).


\bibitem{esaki2}
T. Jiang, K. Chang, L.-M. Si, L. Ran, and H. Xin, Active Microwave Negative-Index Metamaterial Transmission Line with Gain, Phys.
Rev. Lett. 107, 205503 (2011).

\end{thebibliography}
\end{document}